\documentclass[11pt,a4paper]{article}
\usepackage{txfonts}
\usepackage{graphicx}
\usepackage[top=1.5cm,left=1.8cm,right=1.8cm,bottom=1.5cm,headsep=7mm,footskip=7mm]{geometry}
\usepackage{fancyhdr}
\usepackage{multicol}
\usepackage[leftcaption]{sidecap}
\usepackage{framed}
\usepackage[font=it,labelfont=bf]{caption}
\usepackage{url}
\usepackage[usenames,dvipsnames]{color}
\usepackage{multicol}
\usepackage{eurosym}
\usepackage{natbib}
   {\vspace{-3mm}
    \begin{list}{$\bullet$}{%
       \leftmargin 7mm
       \rightmargin 7mm
       \parsep 0mm
       \itemsep 0mm
       \listparindent 0mm
       \itemindent 0mm
       \labelsep 2mm
       \itemindent 0mm
      }
   }
   {\end{list}
}
\newenvironment{packed_item}{
  \begin{itemize}
    \setlength{\itemsep}{1pt}
    \setlength{\parskip}{0pt}
    \setlength{\parsep}{0pt}
  }{\end{itemize}}

\begin{document}

\vspace{1cm}
\begin{center}
  \Large
  {\bf Monitoring young associations and open clusters \\
    with Kepler in two-wheel mode} \\
  \vspace{0.5cm}
  \large
  S.~Aigrain$^{(1)}$, S.~Alencar$^{(2)}$, R.~Angus$^{(1)}$,
  J.~Bouvier$^{(3)}$, E.~Flaccomio$^{(4)}$, 
  E.~Gillen$^{(1)}$, J.~Guzik$^{(5)}$, L.~Hebb$^{(6)}$,
  S. Hodgkin$^{7)}$, 
  A.~McQuillan$^{(8)}$, G.~Micela$^{(4)}$, E.~Moraux$^{(3)}$, H.~Parviainen$^{(1)}$,
  S.~Randich$^{(9)}$, S.~Reece$^{(1)}$, S.~Roberts$^{(1)}$, K.~Zwintz$^{(11)}$ \\
\end{center}
{\small 
\begin{tabular}{p{0.5\linewidth}p{0.5\linewidth}}
\noindent $^{(1)}$ University of Oxford, United
Kingdom (UK) &
\noindent $^{(2)}$ Departamento de F{\'i}sica--ICEx, UFMG, Brazil \\
\noindent $^{(3)}$ IPAG,
Grenoble, France &
\noindent $^{(4)}$ Osservatorio Astronomico di Palermo, Italy \\
\noindent $^{(5)}$ Los Alamos National Laboratory, NM &
\noindent $^{(6)}$ Hobbart \& William Smith Colleges, Geneva, NY \\
\noindent $^{(7)}$ IoA, University of Cambridge,
UK &
\noindent $^{(8)}$ University of Tel Aviv, Israel\\
\noindent $^{(9)}$ Osservatorio Astrophysico di Arcetri,,
Italy &
\noindent $^{(10)}$ Institut f{\"u}r Astrophysik, Universit{\"a}t
Wien, Austria
\end{tabular}}
      
\paragraph{Abstract:} We outline a proposal to
use the Kepler spacecraft in two-wheel mode to monitor a handful of
young associations and open clusters, for a few weeks each. Judging
from the experience of similar projects using ground-based telescopes
and the CoRoT spacecraft, this program would 
transform our understanding of early stellar evolution through the
study of pulsations, rotation, activity, the detection and
characterisation of eclipsing binaries, and the possible detection of
transiting exoplanets. Importantly, Kepler's wide field-of-view would
enable key spatially extended, nearby regions to be monitored in their
entirety for the first time, and the proposed observations would
exploit unique synergies with the GAIA ESO spectroscopic survey and,
in the longer term, the GAIA mission itself.  We also outline possible
strategies for optimising the photometric performance of Kepler in
two-wheel mode by modelling pixel sensitivity variations and other
systematics.

\section{Introduction}
Young associations and open clusters are very useful laboratories to
study star formation and the early stages of stellar evolution, as
they enable us to probe a specific set of properties for a group of
stars sharing the same age and composition but spanning a range of
masses. Time-series photometric observations of such systems, in
particular, can be used to probe a host of important phenomena,
including accretion (for the youngest star forming regions only),
pulsations (from pre- to post-main sequence), rotation and activity.
These observations can also be used to detect and characterise
eclipsing binaries (EBs), and thus to constrain evolutionary models by
measuring the fundamental properties (masses, radii, luminosities and
temperatures) of their component stars in a model-independent
manner. Finally, they can also be used to search for planetary
transits, potentially offering a window into the earliest stages of
the evolution of planetary systems.

Over the past decade, these science goals have motivated a number of
major projects dedicated to the photometric monitoring of star forming
regions and young open clusters. An exhaustive review of these
projects and their achievements would be excessively long, but notable
examples include the Monitor project \citep{aig+07}, which monitored 9
star forming regions and open clusters aged $<200$\,Myr using 2--4\,m
telescopes worldwide. Monitor provided an unprecedented sample of
rotation period measurements for young low-mass stars \citep[see
e.g.][and references therein]{irw+09b,mor+13} and led to the detection
of the lowest mass and youngest stellar eclipsing binary known
\citep[][see Fig.~\protect\ref{fig:mrrel}]{irw+07b}. More recently,
the Palomar Transient Factory Orion project used a 1.2\,m telescope
and focused on the 25\,Ori region, detecting a further 7 pre-main
sequence (PMS) EBs \citep{van+11} as well as what may be the first
exoplanet transiting a PMS star \citep{van+12}.

An important limitation of these projects, however, has been the
time-sampling achievable from the ground. By contrast, the CoRoT and
MOST satellites were able to observe the NGC\,2264 star forming region
continuously for several weeks at a time, first in 2008 and again in
2011/2012, this time as part of a coordinated campaign using Spitzer,
Chandra, VLT-FLAMES, CFHT and a number of other ground-based
telescopes. The data from this program are still being analysed, but
they have already provided fascinating insights into the rapid
evolution of PMS pulsators \citep{gru+12,zwi+13a,zwi+13b} and the
astonishing diversity of the variability of classical T Tauri stars
\citep{ale+10}. The continuous time-sampling enabled a more robust
determination of the rotation period distribution for the cluster
\citep{aff+13}, and led to the detection tens of EBs, including a
dozen which are likely cluster members, one of which is the first
low-mass EB to show evidence of a circumbinary disk \citep{gil+13}.

Unfortunately, NGC\,2264 was the only rich star forming region or open
cluster located within the CoRoT `eyes' (or visibility zone). The
field observed by Kepler during the nominal mission contained only a
handful of moderately old open clusters, although these have already
yielded exciting results on rotation \citep{mei+11}, pulsations
\citep{mig+12} and the frequency of planets around cluster versus
field stars \citep{mei+13}. The possibility of a two-wheel Kepler
mission with decreased photometric precision, but increased
flexibility in terms of pointing, aperture selection and observing
strategy, represents a unique opportunity to monitor other key young
associations and open clusters. A major advantage of Kepler over CoRoT
and previous ground-based programs is its extremely wide
field-of-view. This means that we can observe relatively nearby,
spatially extended regions such as Taurus, the Pleiades, and the
Hyades. Previous projects have avoided them because of the need to
tile observations, but Kepler can observe a significant fraction of
their members in a single shot. The very proximity of these regions
not only makes them cornerstones of our understanding of early stellar
evolution, it also greatly facilitates any follow-up observations.

\section{The target clusters}

\begin{figure}[ht]
  \centering
  \includegraphics[width=0.8\linewidth]{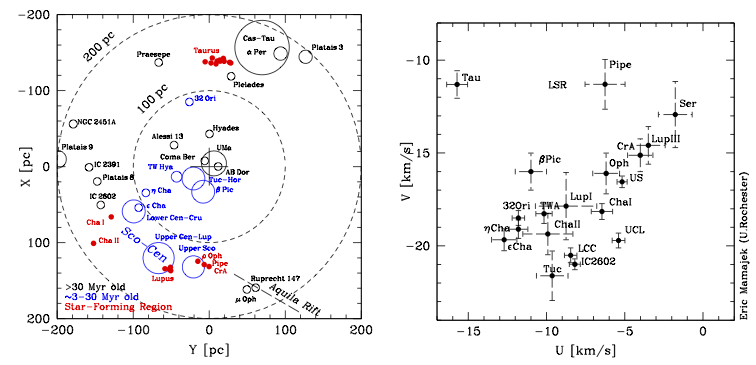}
  \caption{Spatial (left) and kinematic (right) distribution of nearby
    young associations (from
    \protect\citealt{ric+11}).}\label{fig:clusters}
\end{figure}

Figure~\ref{fig:clusters} shows the spatial and kinematic distribution
of young moving groups, associations, and clusters in the solar
neighbourhood, colour-coded according to age. The richest of these
are prime targets for the proposed program, but we also
include a few particularly important clusters at larger distances
and/or older ages.

In the rest of this section, we discuss the most likely targets for
this program, but we stress that this target list is neither
exhaustive nor final, and would need to be adjusted depending on the
total time available, constraints on the satellite pointing,
preliminary indications of the photometric performance of Kepler in
two-wheel mode, and detailed tallies of the latest membership
information for each target.

First, Kepler has the unique ability to observe, in a single shot, the
entire field of most of the nearest and richest star forming regions:
such as Orion \citep{bal08}, $\rho$\,Ophiucus \citep{wil+08},
Chamaeleon \citep{luh08} and Taurus-Auriga \citep{ken+08}
complexes. In Orion alone, we could observe several thousand stars
spanning an age range form 1 to $\sim 12$\,Myr. The
other three are less rich, with a few hundred members each (down to
$V\sim 17$), but even more nearby ($\sim150$\,pc), and their spatially
sparse nature has so far hindered intensive photometric monitoring
campaigns. The two Northern regions are extremely well-studied
already, and the two Southern ones are included in the GES. Combined
with the existing CoRoT observations of NGC\,2264, and with the
infrared variability information collected with Spitzer as part of the
YSOVAR project ({\tt ysovar.ipac.caltech.edu}), Kepler observations of
these three regions would provide fascinating insights into the
evolution of accretion, rotation, activity and structure in the first
few Myr of stellar evolution.

Another very important group of targets are `intermediate age
clusters', ranging from $\sim 30$ to $\sim 150$\,Myr, including for
example: IC\,4665 \citep[][27\,Myr, 350\,pc]{lod+11}, NGC\,2547
\citep[][30\,Myr, 1400\,pc]{jef+04}, $\alpha$\,Per \citep[80\,Myr,
140\,Myr]{lod+12}, Blanco\,1 \citep[][90\,Myr, 130\,pc]{pla+11}, and
the Pleiades \citep[][135\,Myr, 150\,pc]{per+05}.  These clusters will
enable us to probe the transition from PMS to MS and to study stellar
evolution in a lower density environment. $alpha$\,Per and Blanco\,1 in particular
represent a critical age in rotational evolution, and currently have
very little rotation period data. Although some are a little
more distant than the youngest targets, most are still spatially
extended. All those in the Southern hemisphere are in the GES, and the
Northern clusters are extremely well studied already, minimising the
need for dedicated follow-up observations. 

Finally, we include the Hyades\citep[][625\,Myr,47\,pc]{per+98} and
Praesepe \citep[][650\,Myr, 150\,pc]{kra+07}, as `cornerstone' where
FGK stars have all arrived on the MS, while earlier type stars are
already starting to evolve. The Hyades in particular are so near to
the solar system, and so extended, that Kepler really is the only
instrument capable of monitoring them in their
entirety. 

While there are between hundreds to thousands of members in each cluster down
to $V\sim17$ (approximately corresponding to the faint limit of Kepler
in nominal mode), there is a case for pushing to fainter magnitudes,
in order to probe low-mass stars and brown dwarfs. If it is feasible,
a faint cutoff at $V\sim19$ would increase the number of cluster
members monitored size by a factor $\sim 3$ and take us to the hydrogen
burning mass limit for almost all target clusters (well into the
substellar regime for the nearest / youngest).

\section{Science drivers}

\subsection{Pulsations}

\begin{figure}
  \centering
  \includegraphics[height=4cm]{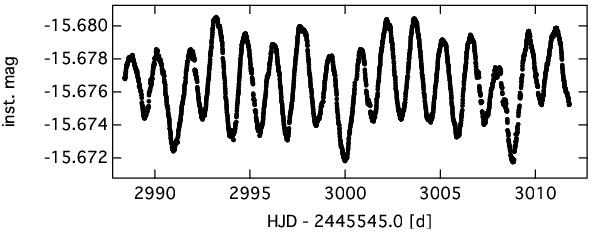} \hfill
  \includegraphics[height=4cm]{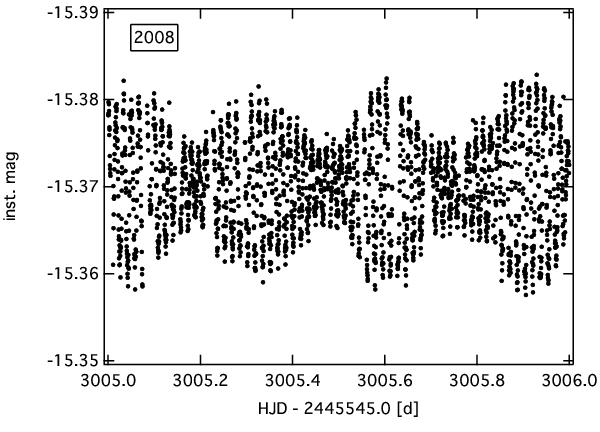}
  \caption{CoRoT light curves of PMS pulsators in NGC\,2264. Left: the
    $\gamma$\,Doradus star
    NGC\,2264~VAS\,20 \protect\citep{zwi+13a}. Right: 1-day section of
   the light curve of $\delta$\,Scuti star HD\,261711 \protect\citep{zwi+13b}.}
  \label{fig:puls}
\end{figure}

Asteroseismology offers a unique window into the interiors of stars,
as amply demonstrated by the very successful asteroseismology program
carried out during the nominal Kepler mission. The reduced photometric
performance expected in two-wheel mode is very likely to preclude the
study of very low-amplitude, Sun-like pulsations, but it should still
be possible to study slower, larger amplitude `classical'
pulsators. It is particularly interesting to study pulsations in young
stars, which are evolving rapidly compared to their older
counterparts, and can thus provide particularly stringent tests of
models of stellar evolution.

Youngs stars with masses between 1.5 and 5\,$M_\odot$ cross the
instability strip in the HR diagram during their evolution towards the
zero-age main sequence, and can thus display $\delta$\,Scuti-like
pulsations. Theory also predicts the existence of other kinds of
pulsators among B to F type PMS and early MS stars, including
$\gamma$\,Doradus stars \citep{bou+11} and slowly pulsating B (SPB)
stars.  Importantly, the interior structure of young stars in this
mass range differs significantly from that of their evolved
counterparts located in the same region of the HR diagram, whereas
their atmospheric properties (and hence colours and spectra) are quite
similar, so asteroseismology uniquely constrains the evolutionary
stage of such stars. Asteroseismology is also, of course, an important
test of theoretical models of the interiors of PMS stars, particularly
when focussing on young open clusters, where all the stars formed from
the same birth cloud. For example, the joint seismic analysis of the 6
pulsators known (at that time) in the 3\,Myr old star forming region
NGC\,2264 by \citet{gue+09} highlighted some important discrepancies
with theoretical predictions.

Pulsating PMS stars have spectral types from B to F, periods ranging
from 30\,min to 1\,day and amplitudes of $\sim 1$\,mmag or less. To
detect and model these pulsations thus requires tight time sampling
(5\,min max) over periods well in excess of a day, as well as a
photometric precision of order 1\,mmag, which is difficult to achieve
from the ground. As a result, much of what we know about these young
pulsators so far has come from a handful of objects located in
NGC\,2264, which has been observed repeatedly by the MOST and CoRoT
satellites, and from dedicated observations of Herbig Ae field stars
with the MOST space telescope.  Notable achievements resulting from
these observations include:
\begin{packed_item}
\item the detection of tens of new PMS $\delta$\,Scuti stars almost
  doubling the total number (Zwintz, priv.\ comm.; see
  Fig.~\ref{fig:puls}, right panel);
\item constraining the evolutionary state of a star from its pulsation
  frequencies \citep{gue+07}; 
\item showing that granulation in the stars’ thin convective envelopes
  might be responsible for the high numbers of low-amplitude
  frequencies observed \citep{zwi+11};
\item the detection of the first PMS star showing hybrid
  $\delta$\,Scuti-$\gamma$\,Doradus pulsations \citep{rip+11};
\item the detection of the first PMS SPB candidates \citep{gru+12}. B-stars have short
  PMS lifetimes, so these objects will enable us to study the
  transition from PMS to MS, i.e.\ from gravitational contraction to
  the onset of hydrogen core burning;
\item the detection of the first PMS $\gamma$\, Doradus candidates
  \cite[][see Fig.~\ref{fig:puls}, left panel]{zwi+13b}. These have
  similar frequencies to SPB stars, so distinguishing between the two
  requires a precise estimate of the effective temperature. The CoRoT
  observations of NGC\,2264 were the first to have the precision and
  time coverage to enable this.
\end{packed_item}
Observing a handful of young clusters and associations with Kepler
would enable us to identify and study these different kinds of
pulsators at a range of ages. We expect to find between 15 and 20
pulsators in each cluster.

\subsection{Accretion, rotation and activity}

\begin{figure}
  \centering
  \includegraphics[height=5cm]{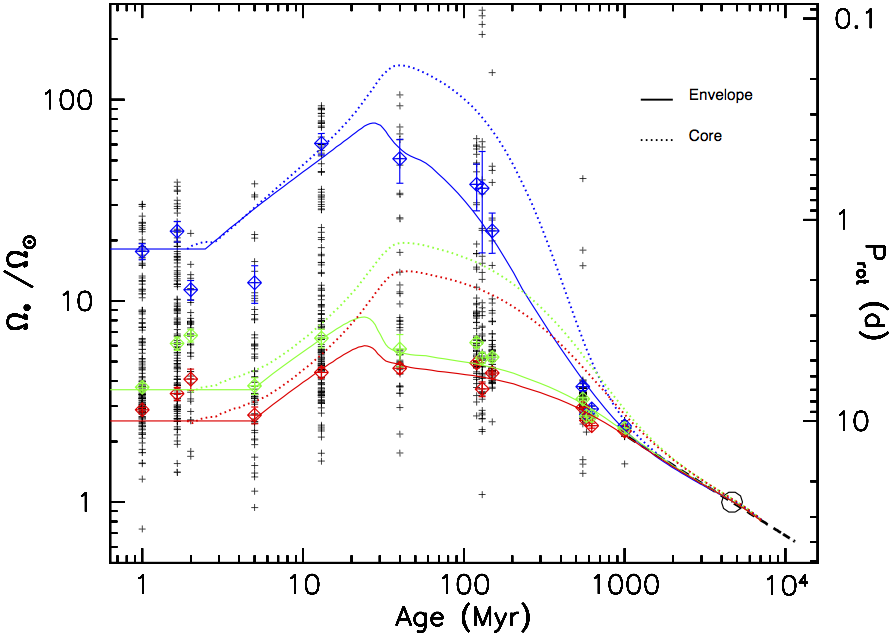} \hfill
  \includegraphics[height=5cm]{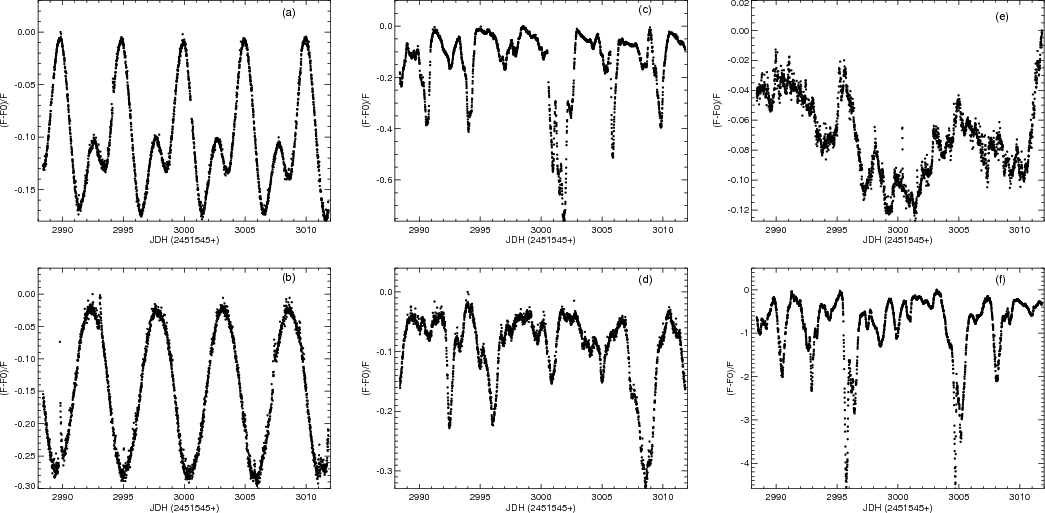} \vspace{1mm}
  \caption{Left: Compilation of rotation period measurements for young
    open clusters with a range of angular evolution models
    \protect\citep{gal+13}. Right: Example light curves of
    spectroscopically identified T Tauri stars observed by CoRoT in
    NGC\,2264\protect\citep{ale+10}, showing spot-like, AA Tau-like
    and irregular variability ($1^{\rm st}$, $2^{\rm nd}$ and 3$^{\rm
      rd}$ column, respectively).}
  \label{fig:rot_ctts}
\end{figure}

The angular momentum evolution of young stars results from a trade-off
between competing effects: contraction onto the main sequence and the
associated spin-up, star-disk interaction (disk-locking), angular
momentum loss via a magnetised wind, and internal re-distribution of
angular momentum as the structure of the star evolves. The past decade
has seen a large increase in the number of rotation period
measurements available for PMS and early MS stars, but theoretical
models still struggle to reproduce all the available data (see
Fig.~\ref{fig:rot_ctts}, left panel). While most
well-studied young clusters and associations have been the subject of
rotation period searches from the ground, these have typically
focussed on the denser areas, and their period sensitivity is far from
uniform. Monitoring selected clusters for a few weeks each would
enable us to complete the census. Typical rotation periods for young
stars range from 1 to 20 days \citep{irw+09b}, and modulation amplitudes from 0.5 to a
few \%, so the main requirement placed by this part of the science
case is the duration of the observations of each region (at least 20
days, ideally up to 40). 

Such observations would also provide an exciting window into the
relationship between the amplitude and period of starspot-related
variability, and -- for the younger associations -- between the latter
and accretion-related variability. The CoRoT observations of NGC\,2264
have shown that about 20\% of the classical T Tauri (CTTS) stars in
the cluster presented light curves that can be attributed to periodic
obscuration of the photosphere by the inner region of the
circumstellar disk (see Fig.~\ref{fig:rot_ctts}, right panel).  These
CTTSs, called AA Tau-like due to their resemblance to the well studied
AA Tau system \citep{bou+99}, offer the unique opportunity to study the
properties of the inner disk region, located at only a few stellar
radii from the star. This also allows the analysis of the dynamical
star-disk interaction.

\subsection{Eclipsing binaries and transits}

\begin{figure}
  \centering   
  \includegraphics[width=0.65\linewidth]{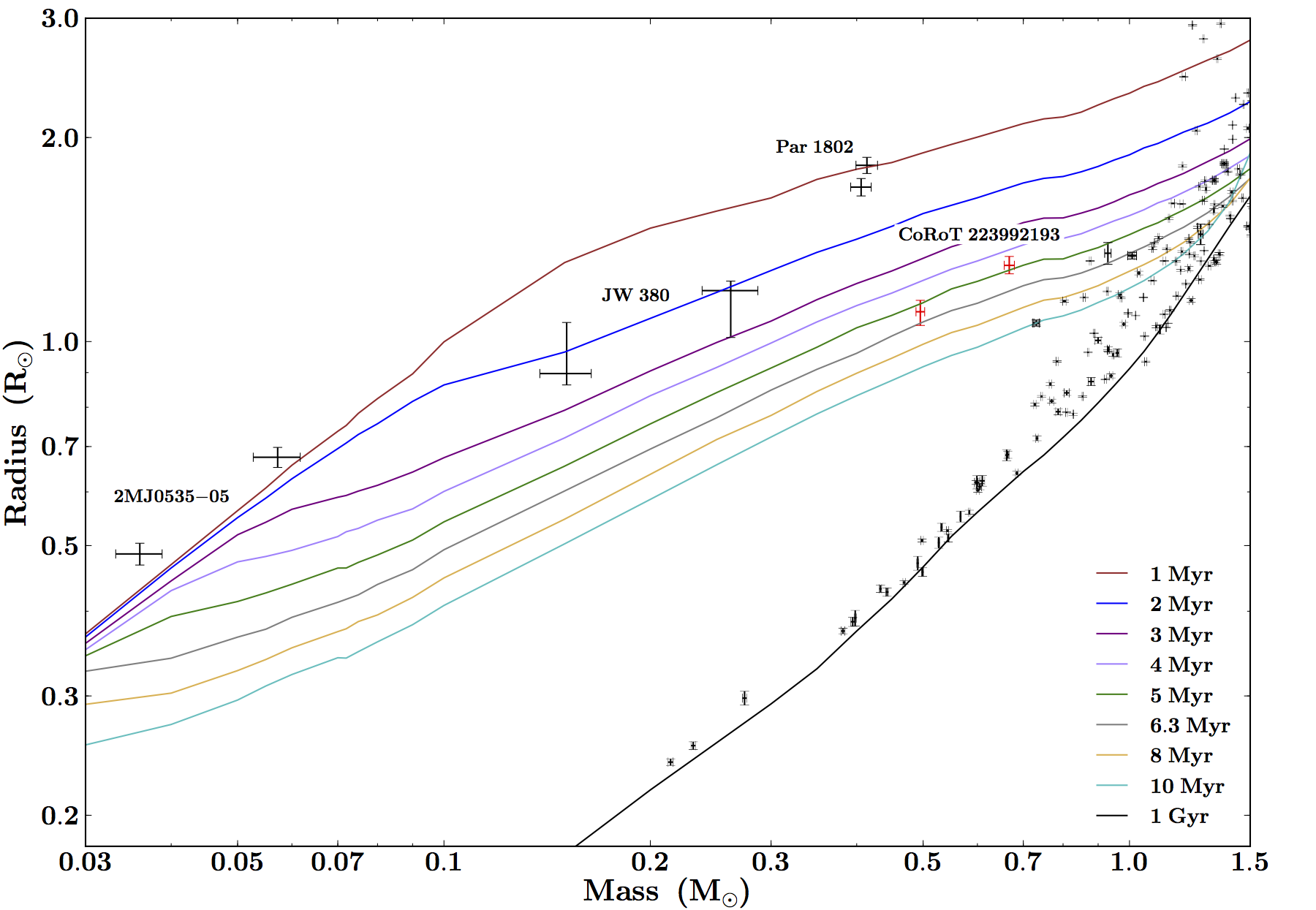}
  \caption{Mass-radius relation for low-mass stars. The lines show,
      from top to bottom, the theoretical isochrones of
      \protect\citet{bar+98}. The black points show measurements for stars with
    masses $< 1.5\,M_{\odot}$ in detached EBs (data from {\tt
      http://www.astro.keele.ac.uk/$\sim$jkt/debdata/debs.html}), with
    one of the new systems discovered by CoRoT in NGC\,2264 shown in
    red \citep{gil+13}. Note the improvement in mass and radius
    determination compared to other PMS systems, due mainly
    to the precise and continuous space-based light curve.}
   \label{fig:mrrel}
\end{figure}

Detached, double-lined eclipsing binaries (EBs) are extremely valuable
objects because their masses, radii, effective temperatures and
luminosities can be determined in a model-independent manner from the
light and radial velocity curves of the system. When these reach a
precision of a few percent or less, they provide one of the most
powerful tests of stellar evolution models available
\citep{and91,tor+10}. As these models underpin most of astrophysics,
it is vital that they are tested as rigorously as possible. The two
components of a given EB can generally be assumed to share the same
age and metallicity, which adds to the tightness of the
constraints. Figure \ref{fig:mrrel} shows the existing mass and radius
measurements for low-mass stars belonging to detached EBs. While there are
now many well-characterised systems on the main sequence, there are
very few on the pre-main sequence (PMS). Furthermore, even in
well-sampled regions of parameter space there are significant
discrepancies between theory and observations, as models tend to
under-predict the radii (or equivalently, over-predict the
temperatures) of low-mass stars. 

\begin{figure}
  \centering   
  \includegraphics[width=0.8\linewidth]{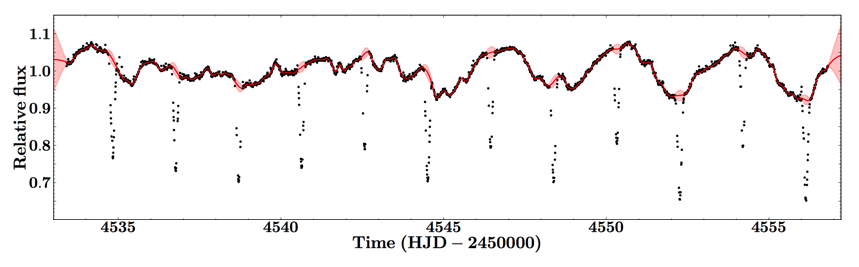} 
  \caption{Light curve of CoRoT 223992193, a low-mass EB discovered by
    CoRoT in NGC\,2264 \citep{gil+13}. The
    out-of-eclipse (OOE) variability is modelled using a Gaussian
    process model (red line).}
   \label{fig:1039}
\end{figure}

Detecting new, young, low-mass EB systems and characterising them in
detail therefore remains a very important goal. The light curve shown
in Figure~\ref{fig:1039} was obtained by CoRoT for a system with
$V=16.8$, and illustrates the kind of precision one can expect to
achieve even in the worst case scenario with Kepler in two-wheel
mode. As shown by the red points in Figure~\ref{fig:mrrel}, this is
sufficient (given suitable follow-up spectroscopic observations to
measure the orbit of the system) to extract useful constraints on the
masses and radii of both components. The continuous sampling
achievable from space also significantly enhances the sensitivity to
moderate period (8--20\,days) EBs, enabling us to test the impact of
varying degrees of mutual interactions between the two stars on their
evolution \citep{cha+07,cou+11,irw+11}.
Based on experience from the Monitor project and the CoRoT
observations of NGC\,2264, we expect to discover around 10 new
eclipsing binaries in each cluster.

Another very exciting prospect is the possible detection of transiting
giant planets on short-period ($<10$\,days) orbits. Detecting
transiting planets in open clusters has proved very difficult from the
ground, not least because the targets tend to be relatively distant,
and thus faint, making spectroscopic follow-up very expensive. The
PTF-Orion project has nonetheless shown that it is possible
\citep{van+12}. Furthermore, the radial velocity detection by
\citet{qui+12} of 2 planets in a sample of 53 stars monitored in
Praesepe indicates that hot Jupiters are at least as common in young
open clusters as in the field. Based on calculations similar to those
performed for Monitor \citep{aig+07}, we expect between 0 and 2
transiting planets to be detectable in each cluster; the exact number
is very sensitive to the photometric precision (not yet known) and time coverage.

\section{Synergy with other projects}

Aside from the aforementioned CoRoT and MOST observations, the
proposed program is similar to the Monitor \citep{aig+07}, but would
supersede it significantly in terms of field-of-view, precision and
sampling. There is also some overlap with the science goals of YSOVAR project
({\tt http://ysovar.ipac.caltech.edu/}),
which monitored a number of star forming regions with warm
Spitzer. The wavelength
range is clearly complementary, and again Kepler observations would
provide improved sampling over a much wider field of
view. Re-observing some of the targets already observed by Monitor and
YSOVAR will be valuable per se, for example to chart evolution in disk
and star-spot activity in individual stars on multi-year timescales.

Importantly, the proposed observations will benefit from a very natural
synergy with the Gaia-ESO Public Spectroscopic Survey (GES, PIs
G.~Gilmore and S.~Randich, \citealt{gil+12}), a large homogeneous
survey of the distributions of kinematics and chemical element
abundances in the Galaxy, designed to complement the astrometric,
photometric and low-resolution spectroscopic data that GAIA itself will provide.
The GES will obtain spectra of $\sim 100\,000$ stars with VLT/FLAMES,
focussing on well-defined samples of Milky Way field and cluster
stars.  The cluster component of GES will cover 80--90 clusters
spanning a wide range of age, richness, mass, composition, morphology,
etc, ranging from from the closest associations to massive clusters
at few kpc from the Sun, with ages from 1\,Myr to 10\,Gyr. For each
cluster, an unbiased sample of stars are selected in order to derive,
together with the Gaia data, accurate distances, 3-D spatial
distributions and motions. It will also provide precise radial
velocities for each star observed, and hence unbiased (w.r.t.\
activity) estimates of membership probability, as well as the mass,
age, abundance, binarity, lithium abundance, and $v \sin i$ of most
members. For the youngest clusters, GES will discriminate between
classical and weak-lined T Tauri stars (CTTS/WTTS), and thus between
active and inert disks. All of this information will be crucial to
interpret the light curves provided by Kepler. 
The GES observations started in January 2012 and $\sim 15$ clusters
have already been observed, including several of the nearest young
clusters included in this proposal. If the Kepler observations
proposed here go ahead, a detailed plan and formal agreement with the
GES management team will be put in place to make the most of the
synergy, but we note that all GES data
are public in any case.

The GAIA mission itself is due for launch in late 2013 and will, over
the course of a 5 year mission, provide micro-arcsec astrometry,
multi-epoch (70 epochs average, 200 max.) millimag photometry, low-res
(20--50) spectra and, for the brighter objects ($V<15$), moderate
resolution spectra and km/s radial velocities. It will deliver proper
motions with $\sim 20\,\mu$as accuracy down to $V\sim15$, and $\sim
200\,\mu$as accuracy down to $V\sim20$, was well as
sub-milliarcsecond parallaxes down to $V\sim20$\footnote{GAIA
  performance data from {\tt http://www.rssd.esa.int/index.php?page=Science\_Performance\&project=GAIA}.}. This will enable the
derivation of model-independent (kinematic trace-back) ages for young
clusters and associations, and distances with relative precision
better than 10\% down to the brown dwarf regime for the nearest
clusters in this proposal. GAIA will also yield orbital solutions
for multiple systems, including binary stars and gas giant planets at
a few AU from their host star. The first GAIA positions will become
available 2 years after launch and the first parallaxes 6 months after
that, while the full dataset will be delivered at the end of the
mission.

At the preparation stage, preliminary results from GES (and GAIA, depending
on the timing) will make the definition of photometric windows for
individual cluster members to be observed by Kepler extremely
efficient. The detailed characterisation of the clusters and their
members provided by GAIA and GES data will also, of course, facilitate the
interpretation of the Kepler time-series. GAIA will also help in other
way, for example its photometry will extend the
period sensitivity of the Kepler observations; the GES and GAIA RVs
will provide orbital solutions for the EBs identified by Kepler, and
the GAIA astrometric data will complement the EB sample at wide
separations.
%  Another example of the synergy between GAIA and the
% proposed Kepler program is the comparison of disk activity (as
% traced by photometric variability and spectroscopic indicators) and
% kinematics (do strong dynamical interactions in the cluster
% environment lead to disk truncation?).

Finally, we note that many of the stars which we propose to monitor
would be prime targets for follow-up observations with the James Webb
Space Telescope (JWST), due to
their youth and proximity.

\section{Proposed observations}

\subsection{Expected photometric performance}

The reduced pointing performance of Kepler in two-wheel mode is
expected to affect the photometric performance significantly. Based on
simulations performed to date, the call for white papers forecasts a
photometric precision of about 0.5--1\,mmag per 1 minute integration
for a $V=12$ star (cf.\ 30\,ppm during the nominal mission), but warns
that pixel sensitivity variations may limit the overall relative
photometry to 0.3--1\,\%. We take these two extremes as our best and
worst case scenarios, respectively. The best-case scenario might be
attained by implementing novel methods for calibrating the pixel
sensitivity variations, extracting photometry from trailed images,
and/or disentangling the systematic effects from the intrinsic
variability of each star in the light curve itself (see
Section~\ref{sec:phot}). Since the decrease in precision in two- compared to
three-wheel mode is due to systematics, we do not expect the faint-end
performance to be affected significantly, so that $\sim 1\%$ photometry
could be achieved down to Kepler magnitudes of $\sim 19$ (extrapolated
from \citealt{jen+10}). 

\subsection{Observing strategy}

The projected lifetime of Kepler in two-wheel mode is 1 to 2 years. In
a one-year program, we would be able to monitor 8-12 clusters for 4 to
6 weeks each.  The main driver for the duration of the observations is
sensitivity to longer rotation periods and longer period EBs, but the
duration will also affect -- for example -- the precision of the
asteroseismic analysis. Given two years, we could extend the target
set, or the duration of each run (increasing sensitivity to
long-period rotators and EBs) or return to some of the clusters
observed in year 1 to probe long-term evolution of the variability
properties. Even if the time available is much more restricted, so
that only a few clusters can be observed, this will already represent
a many-fold increase in the number and range of stars monitored in
this way.

Target lists for each cluster will be constructed by collating all the
membership information available in the literature and from the GES
and GAIA. The number of known members in each of the clusters listed
in the previous section ranges from a few hundred to several thousand.
Photometric apertures will be defined so as to follow the trail of
each star, and will be allocated in priority to known members of the
cluster. The remaining telemetry can be used to observe other targets
in the same fields. In the denser, central regions of some clusters,
it may be advantageous to download contiguous sections of the detector
by collating multiple apertures.

The standard 30 min cadence is acceptable for some of the science
goals discussed above (rotation, activity, EBs and transits), but some
require a cadence of $\sim 5$\,min or better (pulsations, rapid
variability in T Tauri stars). Whether a subset of the targets are
selected to be observed at the standard short cadence (1\,min), or a
different combination of exposure times, it is clear that the time
sampling requirements of this program are not expected to be
problematic.  While observing any given region once would already
represent a significant advance, if the possibility arises, it would
also be interesting to revisit one or more of the targets after one or
more years, as done with CoRoT for NGC\,2264, to track secular changes
in the different types of variability being studied.

As discussed previously, a lot of information is already available
about the properties of the target regions and their
members. Nonetheless, if this program goes ahead, we will also seek to
organise simultaneous monitoring campaign with other ground- and
space-based facilities (spanning complementary wavelength ranges), as
we have done in the past for NGC\,2264.

\section{Possible strategies for optimising photometric
  performance}
\label{sec:phot}

The reduced pointing performance affects the photometric performance
in two ways: through pixel sensitivity variations (the star samples
many more pixels during an observation, each of which may have a
slightly different sensitivity) and because the images will be come
trailed for any integrations longer than about 5 min.

\subsection{Modelling inter- and intra-pixel sensitivity variations}

The best way to reduce the impact of inter- and intra-pixel variations
may be to devise a novel way of calibrating them prior to the
observations. In the absence of such a development, however, it might
be possible to calibrate them, on a star-by-star basis, from the pixel
time-series themselves. Below we outline a simple model for doing
this. This model relies on a number of simplifying assumptions, some
of which may well be excessively naive, but we merely suggest it here
as an idea. We have not had the opportunity to implement and test it
yet, but we would be interested in
working with the science office to do so, if the opportunity arises.

Consider one star whose flux and position on the detector at time $t$ are given by
$S(t)$, $x_0(t)$ and $y_0(t)$, respectively. The ultimate quantity of
interest is $S$, which is not known \emph{a priori}. On the other
hand, it is reasonable to assume that $x_0$ and $y_0$ are known (from
individual centroid measurements and/or global modelling of the
satellite pointing). The spatial distribution of the flux on the
detector is defined by the point-spread function, $P(\delta x,\delta
y)$, where $\delta x$ and $\delta y$ are the departures from the
star's nominal position in the $x$ and $y$ directions,
respectively. For now we assume that the point-spread function for a
given star is constant in time -- we address the time-dependence of
the PSF introduced by the pointing drift later.  Again,
it is reasonable to assume that the PSF is well-known, or at least
that it can be reduced to a known function with a small number of free
parameters. The flux recorded during the $k^{\rm th}$ integration by
the $(i,j)^{\rm th}$ pixel is then
\begin{equation}
\label{eq1}
F_{ijk} = S(t_k) \, R_{ij} \, \int_{x=i-0.5}^{x=i+0.5}
\int_{y=j-0.5}^{y=j+0.5} D(x-i,y-j) \, P(x-x_0(t_k),y-y_0(t_k)) \,
{\rm d}x \, {\rm d}y,
\end{equation}
where $R_{ij}$ is the (unknown) peak sensitivity of the $(i,j)^{\rm
  th}$ pixel, which we assume to be constant in time, and $D(\delta
x,\delta y)$ represents the relative intra-pixel sensitivity
variations, where $D=1$ for $\delta x = \delta y = 0$. Once more, we
have assumed that $D$ is independent of time, that it is the
same for all pixels, and that it can be described by a simple
parametric function (in the simplest extreme, $D=1$ everywhere).

If there are $N$ observations spanning $M$ pixels, i.e.\ $N \times M$
data points in the entire time-series, the above model has $N+M+K$
free parameters, where $K$ is the number of parameters associated with
the functions $P$ and $D$, and is assumed to be small. In practice,
the effective number of data points will be smaller, as only $M'<M$
pixels will contribute significantly to the PSF at any given time. On
the other hand, as the pointing of the satellite will be reset
periodically, the same pixels will be sampled multiple
times. Therefore, overall the problem should still be well
constrained. Where appropriate, additional leverage may also be gained
by placing certain restrictions on the form of $S$ (e.g. by
constraining it to vary smoothly, quasi-periodically, etc\ldots), if
the star in question is a known type of variable, for example.

The practical implementation of this model will be
challenging, due to the large number of parameters. However, we 
do expect it to be feasible, for example using advanced Markov Chain
Monte Carlo sampling methods \citep[see e.g.][]{for+13} specifically designed to
explore large and complex parameter spaces. Whatever the inference
method used, this will be a computationally intensive process, and it
may be that this approach could
be applied only to specific objects where attaining maximum precision
is particularly important. 
 
\paragraph{Moving stars and trailed images}

In practice, the pointing drift will cause the PSF to be elongated for
any integration lasting more than a few minutes. It will also mean
that the position of a star can change significantly, which may alter
the PSF. To address this we replace equation~\ref{eq1} with the
slightly more complex expression:
\begin{equation}
\label{eq2}
F_{ijk} = \int_{t = t_k}^{t=t_k+\delta t} S(t_k) \, R_{ij} \, \int_{x=i-0.5}^{x=i+0.5}
\int_{y=j-0.5}^{y=j+0.5} D(x-i,y-j) \, P(x-x_0(t),y-y_0(t),x_0(t),y_0(t)) \,
{\rm d}x \, {\rm d}y \, {\rm d}t,
\end{equation}
where $P$ is now a function of the instantaneous position of the star as well as 
the departure from this position. 

% \smallskip

% A potentially important caveat to mention is the spatially variable
% background to be expected in the youngest clusters due to nebulosity. 

\subsection{\emph{A posteriori} correction of systematic effects}

In this section we discuss potential strategies for mitigating the
effects of any instrumental systematics, which are not calibrated out
at the light curve extraction stage.
During the nominal mission, the light curves extracted and calibrated
by the standard pipeline displayed systematic effects, which were
corrected in part by the pre-search data conditioning (PDC) step,
albeit often at the expense of the intrinsic variability
(other than transits). 

\begin{figure}
  \centering   
  \includegraphics[width=0.48\linewidth]{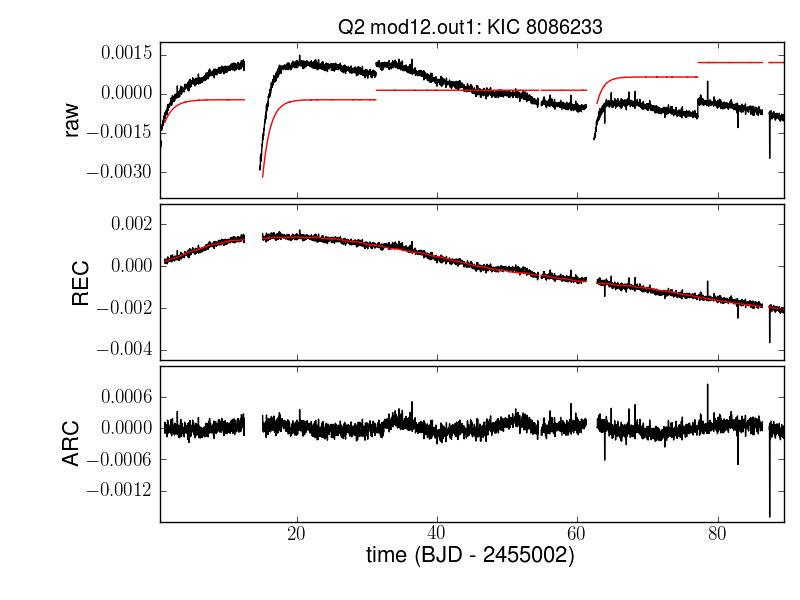} \hfill
  \includegraphics[width=0.48\linewidth]{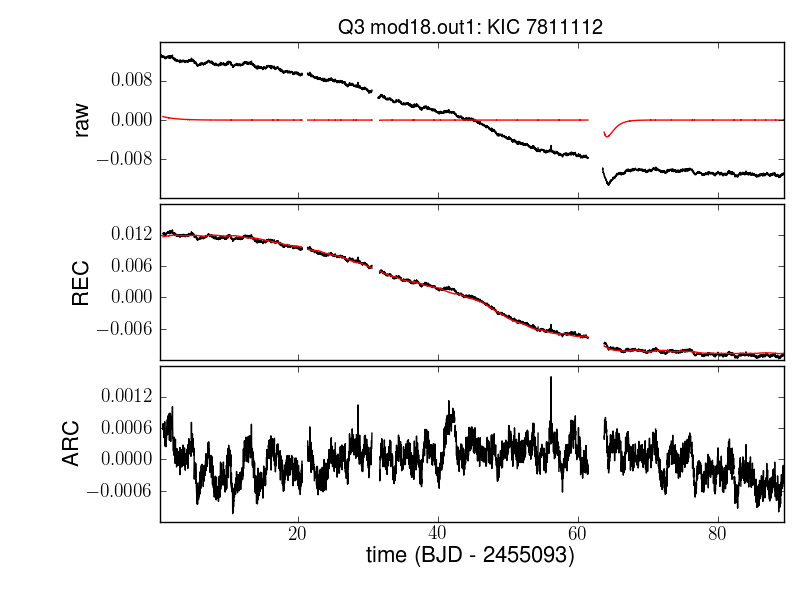}
   \caption{Two-step systematics correction for two representative
     examples from quarters 2 (left) and 3 (right). Top: raw data
     (black) and correction applied for discontinuities (`REC', red). Middle: REC-corrected data (black) and
     correction applied for common-mode systematic trends (ARC,
     red). Bottom: ARC-corrected data.}
   \label{fig:sys}
\end{figure}

To address this problem, we have adopted a two-step approach, which is
still under development but is giving good results. We model
common-mode systematic trends by modelling each light curve in turn as
a linear combination of all the other light curves, and then applying
a statistical entropy criterion to ensure that any trends identified
in this manner are genuinely systematic \citep{rob+13}. This
algorithm, which we call `ARC' (Astrophysically Robust Correction of
systematic trends), uses a Bayesian approach with shrinkage priors to
avoid overfitting, which we implement within a variational
inference framework to ensure computational efficiency. On the other
hand, some instrumental effects, in particular the discontinuities and
thermal decays associated with monthly data download events, are
present in all the light curves, but cannot be represented adequately
by a linear basis model such as the one used by the ARC. We model
these on a star by star basis, postulating a functional form for the
systematic effect, and modelling it at the same time as the stellar
variability itself, which we treat as a Gaussian process (GP, a very
flexible, yet robust class of models, where functions are parametrised
indirectly through their covariance properties). We refer to this as
fault rectification (REC), and apply it before the ARC (see
Figure~\ref{fig:sys} for examples).
 
In two-wheel mode, we anticipate that common-mode systematics will be
less widespread, as inter-pixel sensitivity variations (rather than
global effects such as focus changes associated with the thermal
relaxation of the satellite) are expected to dominate the systematics
budget. Therefore, it may not be possible to describe any systematic effects
which make it through the light curve as a linear combination of a
small number of basis trends common to many light curves, which is the
basis of the ARC algorithm. Indeed, the form of the systematics might well be
unique to each light curve. On the other hand, it is also unlikely
that we will be able to describe them using a specific functional
form, as we have done for the thermal decay events during the nominal
mission. 
However, the systematics are likely to be correlated in some way with
the position of the star on the detector. We therefore suggest that it
might be possible to model them using a GP, whose inputs are the $x$
and $y$ positions of the star on the detector. This is very similar to
the technique developed by our group to treat systematics in Hubble
Space Telescope exoplanet transmission spectra observations
\citep{gib+12}, which we have also successfully used to model the
pixel response function of Spitzer (Evans
et al., in prep.).

\section{Conclusions}

We have shown that Kepler in two wheel mode could be a very powerful
tool to monitor nearby, spatially extended young clusters and
associations, thanks to its unique field of view, continuous coverage,
and a photometric performance which, while reduced, is still likely to
be very good. For example, observing 8--12 clusters for 4--6 weeks
each, would lead to an order of magnitude increase in the number of
known PMS and early MS pulsators, and enable us to chart the evolution
of intermediate and low-mass stars onto the main sequence in
unprecedented detail. It would complete enable us to probe the full
diversity of accretion and activity-induced variability right (right
into the brown dwarf regime for the some clusters), and complete the
census of rotation periods in some of the nearest and best studied
young associations. Finally, it will also lead to an order of
magnitude increase in the number of well-characterised, young
eclipsing binaries, and may lead to the detection of a few young
transiting planets. We have shown that there are valuable synergies
between the proposed observations and the GAIA-ESO survey, as well as
the GAIA space mission itself. Finally, we have also outlined possible
strategies for optimising the photometric performance of Kepler in
two-wheel mode, which could be used for a wide range of observing
programs.


\begin{thebibliography}{99}
\begin{multicols}{2}
  {\footnotesize \bibitem[\protect\citeauthoryear{Affer et
      al.}{2013}]{aff+13} Affer L., et al., 2013, MNRAS, submitted
    \vspace{-3mm} % NGC2264 rotation
  \bibitem[\protect\citeauthoryear{Aigrain et al.}{2007}]{aig+07}
    Aigrain S., et al., 2007, MNRAS, 375, 29
    \vspace{-3mm} % Monitor intro
  \bibitem[\protect\citeauthoryear{Alencar et al.}{2010}]{ale+10}
    Alencar S.~H.~P., et al., 2010, A\&A, 519, A88
    \vspace{-3mm} % NGC2264 accretion
  \bibitem[\protect\citeauthoryear{Andersen}{1991}]{and91} Andersen
    J., 1991, A\&ARv, 3, 91 \vspace{-3mm} % masses radii normal stars
  \bibitem[\protect\citeauthoryear{Bally}{2008}]{bal08}
    Bally, J., 2008, Handbook of Star Forming Regions, Vol.\ 1
    \vspace{-3mm} % ONC
  \bibitem[\protect\citeauthoryear{Baraffe et al.}{1998}]{bar+98}
    Baraffe I., et al., 1998, A\&A,
    337, 403 \vspace{-3mm} % evolutionary models
  \bibitem[\protect\citeauthoryear{Bouabid et
      al.}{2011}]{bou+11} Bouabid M.-P., et al., 2011,
    A\&A, 531, A145 \vspace{-3mm} % Theory PMS gamma dor
  \bibitem[\protect\citeauthoryear{Bouvier et al.}{1999}]{bou+99}
    Bouvier J., et al., 1999, A\&A, 349, 619 \vspace{-3mm} % AA TAU
  \bibitem[\protect\citeauthoryear{Bouvier}{2009}]{bou09} Bouvier J.,
    2009, EAS, 39, 199 \vspace{-3mm} % rotation review
  \bibitem[\protect\citeauthoryear{Chabrier, Gallardo, \&
      Baraffe}{2007}]{cha+07} Chabrier G., Gallardo J., Baraffe I.,
    2007, A\&A, 472, L17
    \vspace{-3mm} % evolution low-m S and BD binaries
  \bibitem[\protect\citeauthoryear{Coughlin et al.}{2011}]{cou+11}
    Coughlin J.~L., et al., 2011, AJ, 141, 78
    \vspace{-3mm} % Kepler low-mass EBs
  \bibitem[\protect\citeauthoryear{Foreman-Mackey et
      al.}{2013}]{for+13} Foreman-Mackey D., et al., 2013, PASP, 125,
    306 \vspace{-3mm} % emcee
  \bibitem[\protect\citeauthoryear{Gallet 
      \& Bouvier}{2013}]{gal+13} Gallet F., Bouvier J., 2013, A\&A,
    556, A36 \vspace{-3mm} % rotation models 
  \bibitem[\protect\citeauthoryear{Gibson et al.}{2012}]{gib+12}
    Gibson N.~P., et al., 2012, MNRAS, 419, 2683 \vspace{-3mm} % GP systematics
  \bibitem[\protect\citeauthoryear{Gillen et al.}{2013}]{gil+13}
    Gillen, E., et al., 2013, A\&A, submitted
    \vspace{-3mm} % NGC2264 EB1039
  \bibitem[\protect\citeauthoryear{Gilmore et al.}{2012}]{gil+12}
    Gilmore G., et al., 2012, Msngr, 147, 25
    \vspace{-3mm} % GAIA ESO survey
\bibitem[\protect\citeauthoryear{Gruber et
      al.}{2012}]{gru+12} Gruber D., et al., 2012, MNRAS, 420, 291
    \vspace{-3mm} % NGC2264 SPB candidates
  \bibitem[\protect\citeauthoryear{Guenther et
      al.}{2007}]{gue+07} Guenther D.~B., et al., 2007, ApJ, 671,
    581\vspace{-3mm} % constraining evol state
  \bibitem[\protect\citeauthoryear{Guenther et al.}{2009}]{gue+09}
    Guenther D.~B., et al., 2009, ApJ, 704, 1710
    \vspace{-3mm} % ensemble astero NGC2264
  % \bibitem[\protect\citeauthoryear{Irwin et al.}{2006}]{irw+06} Irwin
  %   J., Aigrain S., Hodgkin S., Irwin M., Bouvier J., Clarke C., Hebb
  %   L., Moraux E., 2006, MNRAS, 370, 954
  %   \vspace{-3mm} % Monitor rotation M34
  % \bibitem[\protect\citeauthoryear{Irwin et
  %     al.}{2007a}]{irw+07a} Irwin J., Hodgkin S., Aigrain S., Hebb L.,
  %   Bouvier J., Clarke C., Moraux E., Bramich D.~M., 2007, MNRAS, 377,
  %   741 \vspace{-3mm} % Monitor rotation NGC2517
   \bibitem[\protect\citeauthoryear{Irwin et
       al.}{2007b}]{irw+07b} Irwin J., et al., 2007, MNRAS, 380, 541
    \vspace{-3mm} % Monitor EB JW380
  %\bibitem[\protect\citeauthoryear{Irwin et
   %   al.}{2008a}]{irw+08a} Irwin J., Hodgkin S., Aigrain S., Bouvier
    %J., Hebb L., Moraux E., 2008, MNRAS, 383, 1588
   %  \vspace{-3mm} % Monitor rotation NGC2547
  % \bibitem[\protect\citeauthoryear{Irwin et
  %     al.}{2008b}]{irw+08b} Irwin J., Hodgkin S., Aigrain S., Bouvier
  %   J., Hebb L., Irwin M., Moraux E., 2008, MNRAS, 384, 675
  %   \vspace{-3mm} % Monitor rotation NGC2362
  % \bibitem[\protect\citeauthoryear{Irwin et
  %     al.}{2009a}]{irw+09a} Irwin J., Aigrain S., Bouvier J., Hebb L.,
  %   Hodgkin S., Irwin M., Moraux E., 2009, MNRAS, 392, 1456
  %   \vspace{-3mm} % Monitor rotation M50
  \bibitem[\protect\citeauthoryear{Irwin \& Bouvier}{2009b}]{irw+09b}
     Irwin J., Bouvier J., 2009, IAUS 258
    \vspace{-3mm} % Monitor rotation review
  \bibitem[\protect\citeauthoryear{Irwin et al.}{2011}]{irw+11} Irwin
    J.~M., et al., 2011, ApJ, 742, 123 \vspace{-3mm} % long period EB
  \bibitem[\protect\citeauthoryear{Jeffries et
      al.}{2004}]{jef+04} Jeffries R.~D., et al., 2004, MNRAS, 351,
    1401 \vspace{-3mm} % NGC2547
  \bibitem[\protect\citeauthoryear{Jenkins et 
      al.}{2010}]{jen+10} Jenkins J.~M., et al., 2010, ApJ, 713, 
    L120 % phot perf Kepler
  \bibitem[\protect\citeauthoryear{Kenyon et al.}{2008}]{ken+08}
    Kenyon S.\,J., G{\'o}mez M., Whitney B.\,A., 2008, Handbook of
    Star Forming Regions, Vol. 1 \vspace{-3mm} % Taurus-Auriga
  \bibitem[\protect\citeauthoryear{Kraus \&
      Hillenbrand}{2007}]{kra+07} Kraus A.~L., Hillenbrand L.~A.,
    2007, AJ, 134, 2340 \vspace{-3mm} % Praesepe
  \bibitem[\protect\citeauthoryear{Lodieu et al.}{2011}]{lod+11}
    Lodieu N., et al., 2011, A\&A, 532, A103  \vspace{-3mm} % IC4665
  \bibitem[\protect\citeauthoryear{Lodieu et
      al.}{2012}]{lod+12} Lodieu N., et al., 2012, MNRAS,
    426, 3403  \vspace{-3mm} % alpha Per
  % \bibitem[\protect\citeauthoryear{L{\'o}pez Mart{\'{\i}} et
  %     al.}{2013}]{lop+13} L{\'o}pez Mart{\'{\i}} et al., 2013, A\&A,
  %   556, A144 \vspace{-3mm} % chamI
  \bibitem[\protect\citeauthoryear{Luhman}{2008}]{luh08} Luhman, K.,
    2008, Handbook of Star Forming Regions, Vol.\ 2 \vspace{-3mm} % Chamaeleon
  \bibitem[\protect\citeauthoryear{Meibom et
      al.}{2011}]{mei+11} Meibom S., et al., 2011, ApJ, 733, L9
    \vspace{-3mm} % Kepler cluster study
  \bibitem[\protect\citeauthoryear{Meibom et
      al.}{2013}]{mei+13} Meibom S., et al., 2013, Natur, 499, 55
    \vspace{-3mm} % Kepler planet cluster field
  \bibitem[\protect\citeauthoryear{Miglio et
      al.}{2012}]{mig+12} Miglio A., et al., 2012, MNRAS, 419, 2077
    \vspace{-3mm} % Kepler astero old OC
  \bibitem[\protect\citeauthoryear{Moraux et
      al.}{2013}]{mor+13} Moraux E., et al., 2013, A\&A in press,
    arXiv:1306.6351 \vspace{-3mm} % Monitor rotation h Per
  \bibitem[\protect\citeauthoryear{Platais et al.}{2011}]{pla+11}
    Platais I., et al., 2011, MNRAS, 413, 1024 \vspace{-3mm} % NGC2547
  \bibitem[\protect\citeauthoryear{Perryman et al.}{1998}]{per+98}
    Perryman M.~A.~C., et al., 1998, A\&A, 331, 81
    \vspace{-3mm} % Hyades
  \bibitem[\protect\citeauthoryear{Percival et al.}{2005}]{per+05}
    Percival S.~M., Salaris M., Groenewegen M.~A.~T., 2005, A\&A, 429,
    887 \vspace{-3mm} % PLeiades
  \bibitem[\protect\citeauthoryear{Quinn et al.}{2012}]{qui+12} 
    Quinn S.~N., et al., 2012, ApJ, 756, L33 \vspace{-3mm} % RV planets
                                % in praesepe
  \bibitem[\protect\citeauthoryear{Rice et al.}{2011}]{ric+11} 
    Rice E.~L., et al., 2011, ASPC, 448, 481 \vspace{-3mm} % young clusters
  \bibitem[\protect\citeauthoryear{Ripepi et al.}{2011}]{rip+11}
    Ripepi V., et al., 2011, MNRAS, 416, 1535 \vspace{-3mm}% hybrid puls
  \bibitem[\protect\citeauthoryear{Roberts et al.}{2013}]{rob+13}
    Roberts S., et al.., 2013, MNRAS, in
    press, arXiv:1308.3644 \vspace{-3mm} % Kepler systematics ARC
  \bibitem[\protect\citeauthoryear{Torres et al.}{2010}]{tor+10}
    Torres G., Andersen J., Gim{\'e}nez A., 2010, A\&ARv, 18, 67
    \vspace{-3mm} % masses radii normal stars update
  \bibitem[\protect\citeauthoryear{van Eyken et al.}{2011}]{van+11}
    van Eyken J.~C., et al., 2011, AJ, 142, 60
    \vspace{-3mm} % PTF-Orion EBs
  \bibitem[\protect\citeauthoryear{van Eyken et al.}{2012}]{van+12}
    van Eyken J.~C., et al., 2012, ApJ, 755, 42
    \vspace{-3mm} % PTF-Orion planet
  \bibitem[\protect\citeauthoryear{Wilkins et al.}{2008}]{wil+08}
    Wilkins, B.~A. et al., 2008, Handbook of Star Forming Regions,
    Vol.\ 2 \vspace{-3mm} % rho Ophi
  \bibitem[\protect\citeauthoryear{Zwintz}{2008}]{zwi+08}
    Zwintz K., 2008, ApJ, 673, 1088 \vspace{-3mm} % pre-space delta scuti
  % \bibitem[\protect\citeauthoryear{Zwintz et
  %     al.}{2011a}]{zwi+11a} Zwintz K., et al., 2011, A\&A,
  %   533, A133 \vspace{-3mm} 
  \bibitem[\protect\citeauthoryear{Zwintz et al.}{2011}]{zwi+11}
    Zwintz K., et al., 2011, ApJ, 729, 20 \vspace{-3mm} % granulation
  \bibitem[\protect\citeauthoryear{Zwintz et al.}{2013a}]{zwi+13a}
   Zwintz K., et al., 2013, A\&A, 552, A68 \vspace{-3mm} % NGC2264 delta scuti
  \bibitem[\protect\citeauthoryear{Zwintz et al.}{2013b}]{zwi+13b}
    Zwintz K., et al., 2013, A\&A, 550, A121
    \vspace{-3mm} % NGC2264 gamma Dor
  }
\end{multicols}
\end{thebibliography}
\end{document}